\documentclass[preprint]{aastex}
\journalid{}{}
\articleid{}{}
\slugcomment{Astronomical Journal, in press for July 2000}
\shortauthors{Ulvestad}
\shorttitle{Circumnuclear SNRs and \H2 Regions in NGC 253}
\def\H2{\ion{H}{2}}

\begin{document}

\title{Circumnuclear Supernova Remnants and \H2 Regions in NGC 253}

\author{James S.~Ulvestad}
\affil{National Radio Astronomy Observatory\footnote{The
National Radio Astronomy Observatory is a facility of the
National Science Foundation, operated under cooperative
agreement by Associated Universities, Inc.}}
\affil{P.O. Box O, Socorro, NM 87801}
\email{julvesta@nrao.edu}

\begin{abstract}

Archival VLA data has been used to produce arcsecond-resolution
6- and 20-cm images of the region surrounding the nuclear 200-pc
($\sim 15''$) starburst
in NGC~253.  Twenty-two discrete sources stronger than 0.4~mJy 
have been detected within $\sim 2$~kpc ($\sim$3\arcmin) of 
the galaxy nucleus; almost
all these sources must be associated with the galaxy.  
None of the radio sources coincides with a detected X-ray binary, so
they appear to be due to supernova remnants and \H2 regions.  The
region outside the central starburst has a derived radio supernova 
rate of $\lesssim$0.1~yr$^{-1}$, and may account for at least
20\% of the recent star formation in NGC~253.  Most of the newly
identified sources have steep, nonthermal radio spectra, but several 
relatively strong thermal sources also exist, containing the equivalent 
of tens of O5 stars.  These stars are spread over tens of parsecs, and
are embedded in regions having 
average ionized gas densities of $\sim$20--200~cm$^{-3}$, much
lower than in the most active nuclear star-forming regions in
NGC~253 or in the super star clusters seen in other galaxies. 
The strongest region of thermal emission 
coincides with a highly reddened area seen at near-infrared wavelengths,
possibly containing optically obscured \H2 regions.

\end{abstract}


\section{Introduction}

The edge-on spiral galaxy NGC~253 is one of the 
two prototypical, nearby starburst galaxies (along with
M82).   Its distance of 2.5~Mpc \citep{tur85} enables
its starburst disk to be studied at very high linear
resolution, since 1\arcsec\ corresponds to only 12~pc.
The galaxy contains an inner starburst disk roughly
15\arcsec--20\arcsec\ (180--240~pc) in extent that has
been studied in detail, especially by virtue of its
emission in the infrared 
\citep{for91,pin92,ket93,for93,kal94,sam94,bok98,ket99}
and in molecular lines 
\citep{can88,isr95,jac95,pag95,pen96,fra98}.
Subarcsecond-resolution imaging at centimeter wavelengths reveals
at least 64 compact radio sources in the inner disk
\citep{tur85,ant88,ulv91,ulv94,ulv97};
these are thought to be roughly equally divided between
\H2\ regions dominated by thermal emission, and supernova 
remnants dominated by nonthermal emission \citep{ulv97}.  
At the centimeter wavelengths where radio telescopes are most sensitive, 
the poorer resolution and the increasing strength of the more
diffuse galaxy emission lead to confusion that prevents
complete source identification in the inner 200-pc starburst.

The 200-pc region clearly dominates the current star formation
in NGC~253, as shown by the infrared, millimeter, and 
centimeter observations.  Centimeter imaging
of this inner starburst 
indicates a supernova rate of $\leq$0.3~yr$^{-1}$
\citep{ulv97}, which is consistent with results of 
0.1--0.3~yr$^{-1}$ inferred from models of
the infrared emission of the entire galaxy 
\citep{rie80,rie88}.
Near- and mid-infrared imaging at arcsecond resolution
shows a number of emission peaks that do not generally
line up with the radio sources \citep{ket93,sam94,kal94},
while Hubble Space Telescope images show at least four
compact star clusters, which also do not coincide
with individual compact radio sources \citep{wat96}.
Thus, it is apparent that extinction and confusion play 
major roles in the inner disk at different wavebands, 
and it may be that only
the centimeter and millimeter images reveal the regions
of most recent star formation.

Outside the 200-pc starburst disk, NGC~253 may
have a significant amount of star formation,
as shown by the existence of larger-scale radio emission
\citep{hum84,car92,bec94,car96}.
Since this part of the galaxy is much less
affected by confusion and extinction, there is greater
potential for identifying the strongest regions of
star formation and studying them in multiple bands.
However, recent high-resolution studies of the 
galaxy outside the inner few hundred parsecs are uncommon, 
since most of the imaging has been of small fields centered 
on the main starburst disk.  One recent study on the larger scale 
is that of \citet{vog99}, who used ROSAT to identify 73 X-ray 
sources in the bulge, disk, and halo of NGC~253, attributing 
most of them to X-ray binary stars.  To study the population of 
compact radio sources on this large scale, we have
reprocessed the high-resolution 6-cm and 20-cm Very Large Array
(VLA) data obtained in 1987 \citep{ant88,ulv97} in order
to image the entire primary beam of the individual VLA
antennas at high resolution; limited computing resources prevented
this from being done at the time of the observations.  
In this paper, we report
the locations of the compact radio sources outside the
inner 200-pc starburst, compare these locations to images made
in other wavebands, and estimate the supernova
rate outside the inner disk.

\section{Observations and Data Analysis}
\label{obs}

The observations of NGC~253 that we use here
were made in 1987, using the VLA \citep{tho80} in its A
configuration.  Results from imaging the
inner starburst at 6~cm (4860~MHz) were reported by \citet{ant88},
while the 20-cm (1490~MHz) image of the same region was discussed
briefly by \citet{ulv97}. Details of the observations 
were reported in those papers.  Other 6-cm 
observations had much poorer ({\it u,v}\/) coverage than those made
in 1987, so only the 1987 data are considered here.

For self-consistency, the data were re-calibrated and
self-calibrated using the same procedures reported
previously.  We then 
made large-scale ($4096\times 4096$ pixels) images of
the radio emission of NGC~253 to reveal the compact
radio sources seen throughout the primary beam of
the individual 25-m telescopes of the VLA.  These images
covered $26'\times 26'$ at 20~cm and $8'\times 8'$
at 6~cm.  Outside
the starburst disk, confusion was not a significant
issue, but sensitivity was critical.  We therefore 
used natural weighting of the data in the
({\it u,v}\/) plane \citep{sra89,bri99}; this scheme 
produces the maximum sensitivity, at the price of a modest
loss in resolution.  A zero-spacing 
(total) flux density was specified in the imaging,
in order to partially compensate for the lack of short 
interferometer spacings.  This reduced, but did not completely 
remove, systematic effects caused by diffuse emission that was 
not sampled well in our observations.

Table~\ref{tab:obssum} gives the resolution and the r.m.s. noise
achieved for the final images.  
Several effects contribute to reductions in the
sensitivity far from the field center; each is discussed
briefly here.
First, the attenuation of the primary beam of the individual 
VLA antennas reduces the sensitivity by more than 10\% at
6~cm and 20~cm for respective distances from the pointing direction
of greater than $1.\!'7$ and $5.\!'4$.
Corrections for this attenuation were made 
using the AIPS \citep{van96} software developed
by NRAO, and were less than 10\% except for several 
sources (see Section~\ref{background}) far from
the main area of radio emission.  A second effect,
chromatic aberration (``bandwidth smearing'')
due to the non-zero observing bandwidth \citep{brid99},
is more deleterious.  It
reduces the peak flux density by more than a factor of two
at distances greater than 2\arcmin\ from the phase center, and
is progressively worse farther out in the field.  The actual 
phase center of the observations was $\sim 10''$ north of the
apparent galaxy nucleus, at a B1950 position of 
$(\alpha,\delta) = (00^h45^m05.\!^s80,\ -25^\circ33'29.\!''0)$; since
most of the large-scale sources are north of the nucleus, this
actually was slightly beneficial in the final data analysis.
Compared to the chromatic aberration, the reduction in flux
density due to delay smearing \citep{brid99} over the observations' 
10-second averaging period
is negligible.  Finally, the lack of short interferometer spacings 
causes correlated positive or negative emission as high
as 0.08--0.1~mJy over areas of $\sim 50\times 50$ pixels
away from the center of the field.
The final source detection threshold that was used was four times 
the quadrature sum of this systematic value and the r.m.s. noise,
or approximately 0.4~mJy at each wavelength.  However, this
threshold rises substantially at distances more than a few
arcminutes from the phase center.

\placetable{tab:obssum}

\section{Images and Source Identification}
\label{source-id}

\begin{figure}
\figurenum{1}
\plotone{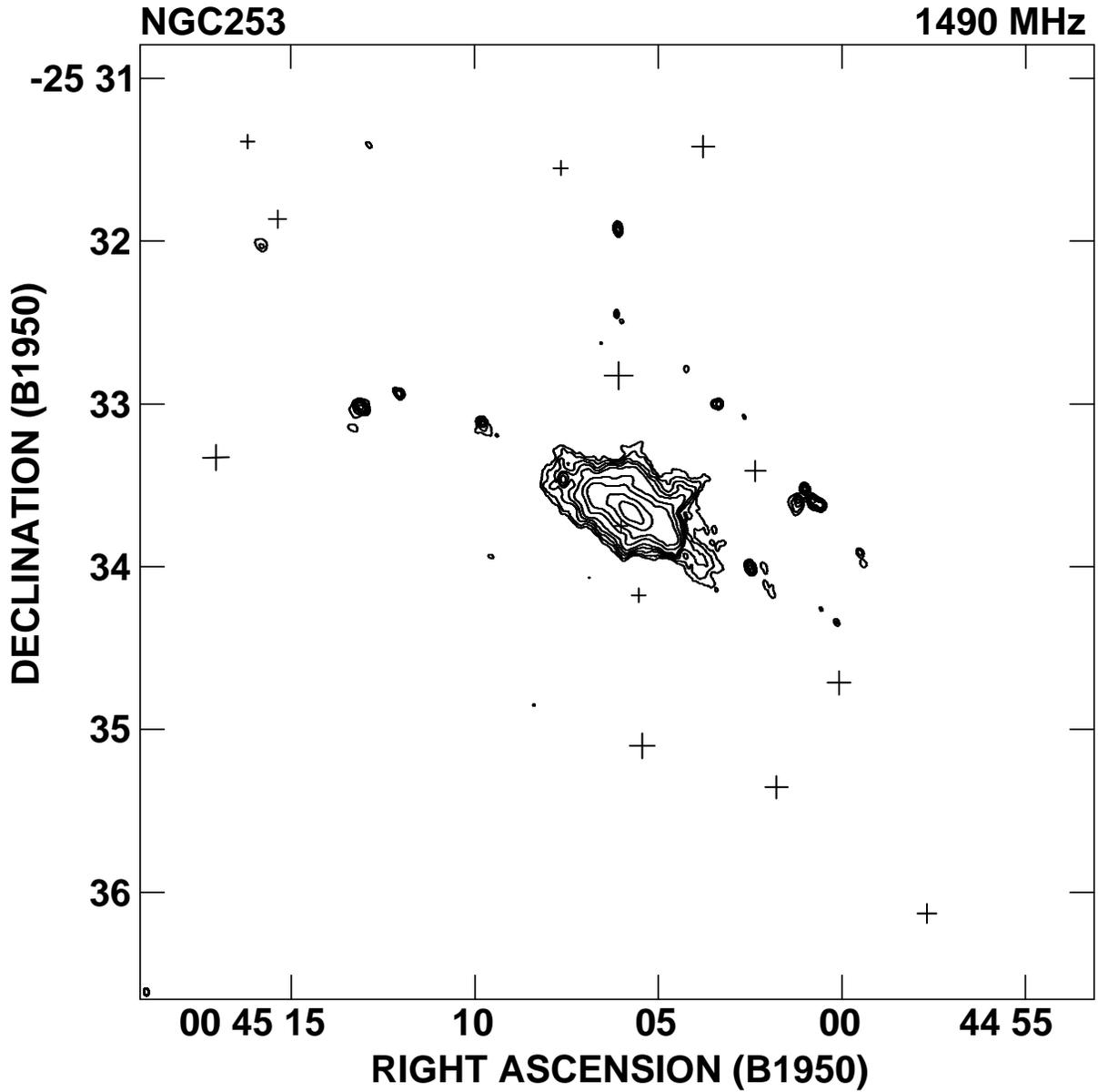}
\caption{Wide-field image of NGC~253 at 20~cm.  Contour
levels are at $-$1, 1, 1.4, 2, 2.8, 4, 8, 16, 64, and 256 times
300~$\mu$Jy~beam$^{-1}$; no negative contours are present in
the image.  The peak flux density is 271~mJy~beam$^{-1}$, and
the restoring beam of $2.\!''77 \times 1.\!''72$
in position angle 7\arcdeg\ is indicated by the minuscule
ellipse in the lower-left corner.  X-ray sources detected by
the ROSAT HRI \citep{vog99} are indicated as crosses, with the
sizes of the crosses showing their positional uncertainty.  \label{lbig}}
\end{figure}

\begin{figure}
\figurenum{2}
\vspace{-2cm}
\plotone{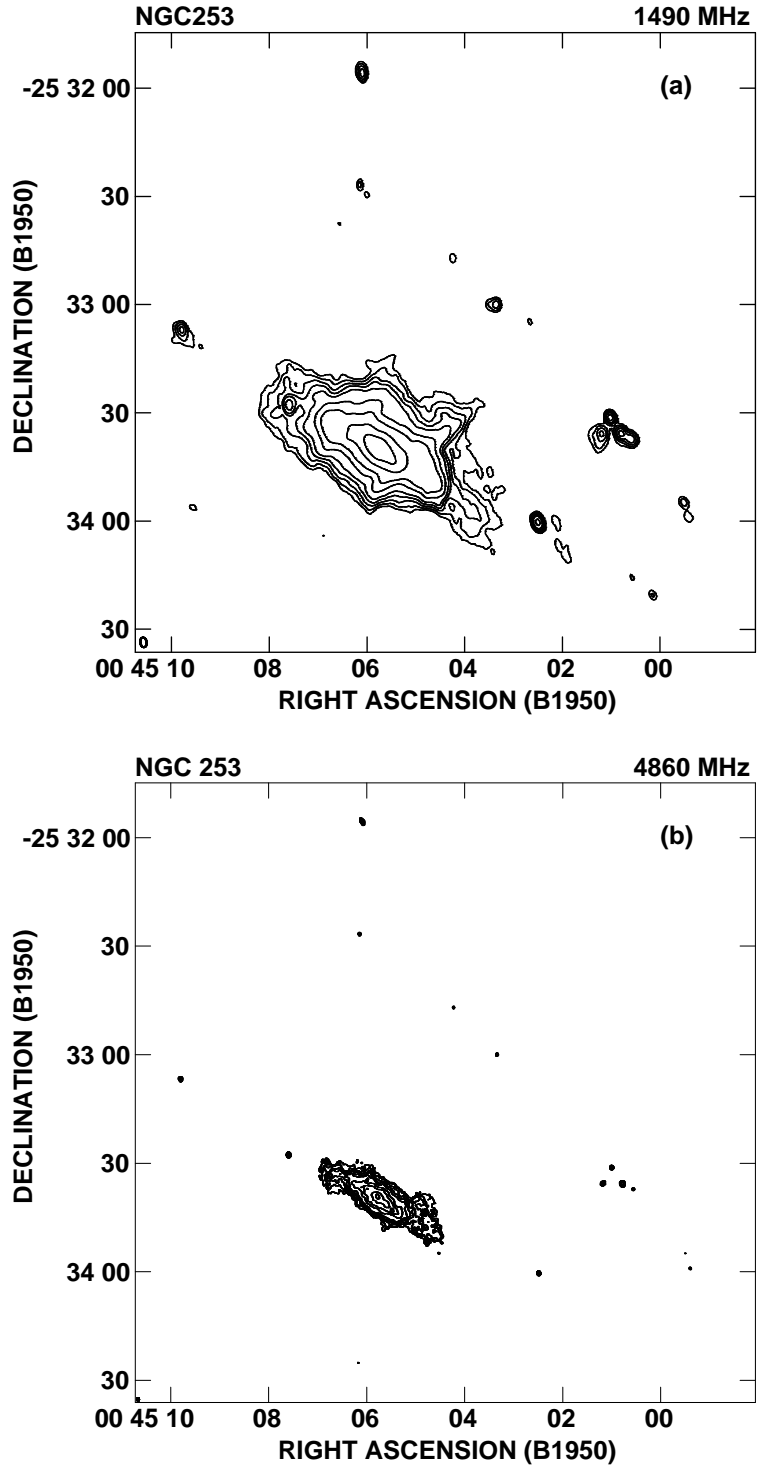}
\caption{Enlargements of 20-cm and 6-cm full-resolution images near the main 
starburst disk of NGC 253.  (a) 20-cm image, with restoring beam and contour 
levels as in Figure~\ref{lbig}.  (b) 6-cm image, with contour levels at 
$-$1, 1, 1.4, 2, 2.8, 4, 8, 16, 64, and 256 times 200~$\mu$Jy~beam$^{-1}$.
The peak flux density is 97~mJy~beam$^{-1}$, and the restoring beam
(lower left) is $0.\!''86 \times 0.\!''51$ in position angle 2\arcdeg.  \label{close}}
\end{figure}

Figure~\ref{lbig} is a portion of the 20-cm image showing the circumnuclear
region of NGC~253 at high resolution.
The locations of compact X-ray
sources detected by \citet{vog99}, using the ROSAT High Resolution 
Imager, are shown as crosses whose sizes indicate
the errors in the X-ray positions.
We have identified all radio sources that have apparent peak flux
densities of at least 0.4~mJy~beam$^{-1}$ at either 6~cm or 20~cm,
outside the central 200-pc starburst.
Figures~\ref{close}a and \ref{close}b are respective enlargements of the 
20-cm and 6-cm images showing
the radio emission from the 200-pc starburst disk and its 
surroundings, including a number
of the circumnuclear compact sources.  Note that the 6-cm image
shows little detail, due to the larger number of beam areas needed 
to cover the given region, but it does serve to illustrate the
relative locations of the detected sources.

\placefigure{lbig}
\placefigure{close}

Table~\ref{tab:sor} lists the source identifications in
the circumnuclear region at the two wavelengths.  (Each source listed
in this table can be seen in the region shown in Figure~\ref{lbig}.)
Included are the source positions, flux
densities at 6~cm and 20~cm, measured sizes, and a few comments.
When compact sources are identified
at both wavelengths, the higher resolution 6-cm position
is quoted; typical position accuracies are
$0.\!''1$--$0.\!''2$.  At 6~cm, flux densities as low as 0.2~mJy are
given for sources that meet the
0.4-mJy threshold at 20~cm, but these flux densities
have very large fractional errors.
  Source flux densities and sizes have been
derived by making Gaussian fits to the individual sources in 
the image plane, and confirmed by integrating over
the region of significant emission.  Most sources are unresolved
in the full-resolution images at both 6~cm and 20~cm, as
indicated by a ``U'' in the table.  The sizes of resolved sources 
are derived from the 6-cm full-resolution image. The total 6-cm flux
densities and approximate sizes found at full resolution generally 
are consistent with values
found from a 6-cm image tapered to the 20-cm resolution.

\placetable{tab:sor}

We estimated flux density errors due to several causes.  First,
the areas of apparent systematic negative or positive 
flux near the main starburst disk, caused by undersampling in
the aperture plane, average as much as
0.08~mJy~beam$^{-1}$ at 6~cm and 
0.10~mJy~beam$^{-1}$ at 20~cm; this offset is denoted by $\sigma_o$.
Second, there is a $\sim 5$\% error in the absolute flux density scales of
each of the two maps, so $\sigma_{\rm sc}=0.05S$, where $S$
is the source flux density.  Third, 
there is a fitting error, $\sigma_f$, caused by
confusion with underlying diffuse emission and uncertainties in the
source fits.  Since the newly
identified sources reported here are outside the main starburst
disk and fairly well isolated, we take the confusion error to be
negligible, so $\sigma_f$ is simply the uncertainty reported by the
least-squares fitting program.  This value includes the
r.m.s. noise given in Table~\ref{tab:obssum}, so the noise
is not included separately.  We combine these errors in
quadrature to get the total flux-density error, $\sigma$, using
\begin{equation}
\sigma^2\ =\ {\sigma_o}^2\ +\ (0.05S)^2\ +\ {\sigma_f}^2\ .
\end{equation}
Thus, the total flux-density error for a 0.4-mJy source 
is typically 0.09~mJy at 6~cm and 0.11~mJy at 20~cm.

\section {Discussion}

\subsection{Are All Detected Sources Associated with NGC~253?}
\label{background}

There are 22 sources above the 0.4-mJy limit at 20~cm, inside a box 
4\arcmin\ (2.8~kpc) on a side and centered on the nucleus of NGC~253.
The density of background sources above this flux-density
threshold is expected to be 
$\sim 10^6$~sr$^{-1}$ \citep{win85}. Therefore, we expect about 
one background source stronger than 0.4~mJy within a field of
16 square arcminutes, implying 
that almost all of the detected sources are associated with NGC~253.
The apparent lack of weak sources at greater distances from the nucleus 
may be caused primarily by chromatic aberration (see Section~\ref{obs});
sources far from the phase center of the observation
have their peak flux densities reduced substantially,
and therefore may fall below the detection threshold.
Observations using narrow spectral-line channels rather than a 
broadband continuum would be needed to determine whether other weak 
compact sources exist farther from the phase center.

Five additional sources were detected at 20~cm within 
an area 12\arcmin\ on a side, centered on NGC~253, and
are listed in Table~\ref{tab:back}.  These are within the
size of the optical galaxy, which has a measured diameter of
25\arcmin\ to 25th magnitude \citep{dev91}.
Images of these sources are
not shown, since they are substantially degraded by
chromatic aberration, which prevents measurement
of the source sizes. Therefore, only the total flux densities are
given in Table~\ref{tab:back}.  The source locations 
are best seen in previous lower-resolution radio images of NGC~253,
particularly that published by \citet{ana90}, though none of
those images resolves source B2 into the three components distinguished 
here.  Only sources much stronger than 0.4~mJy can be detected in
the large region due to the chromatic aberration.  
The expected density of sources 
stronger than 15~mJy at 20~cm is $\sim 10^4$~sr$^{-1}$ \citep{win85},
so we expect to detect less than one background source above this 
threshold.  Source B3 appears to be associated with a ``spur'' 
noted by \citet{car92}, and the three-component source B2
lies along the large-scale galaxy disk, making it probable
that at least B2 and B3 are associated with NGC~253.  

\placetable{tab:back}

\subsection{Circumnuclear Star Formation}

The detected circumnuclear sources (Table~\ref{tab:sor})
outside the central starburst
have typical flux densities of 0.4--3~mJy at 6~cm and 20~cm,
corresponding to radio powers of $3\times 10^{17}$ to
$2\times 10^{18}$~W~Hz$^{-1}$.  Most of them
seem to have relatively steep spectra; for a source with a flux
density of 0.4~mJy at 6~cm and a spectral index of $-0.7$
(with $S_\nu\propto\nu^{+\alpha}$), the total luminosity
between 10~MHz (or 100~MHz) and 100~GHz is 
$\sim 1.1\times 10^{35}$~erg~s$^{-1}$.  This luminosity is
a factor of 100--1000 lower than the luminosities of the
point X-ray sources detected by \citet{vog99}, which they
generally attribute to X-ray binaries.  Since none of the 
radio sources listed in Table~\ref{tab:sor} coincides with a
compact X-ray source, the radio emission is probably not
associated with evolved binaries.

In the main 200-pc starburst disk of NGC~253, \citet{ant88}
deduced that $\sim 100$ compact radio sources exist.
A supernova rate of $\leq 0.3$~yr$^{-1}$ was estimated
from the radio emission \citep{ulv97}, 
while supernova rates of 0.1--0.3~yr$^{-1}$ have been 
derived on other grounds \citep{rie80,rie88}.  Here, we 
report an additional 22 circumnuclear compact radio sources
that lie outside the central 200-pc
starburst, but within 2~kpc (170\arcsec) of the galaxy center.
Since they do not coincide with X-ray binaries, and
most are less than 5--10~pc ($0.\!''4$--$0.\!''8$) in diameter,
they most likely are due to supernova remnants
and \H2 regions, as are the sources in the nuclear region.
The sources can be compared to the galactic supernova
remnant Cas~A.  For an
assumed distance of 2.8~kpc \citep{van71} and a flux density
of $\sim 500$~Jy \citep{baa77}, Cas~A has a
6-cm radio power of $\sim 5\times 10^{17}$~W~Hz$^{-1}$, comparable
to the powers of the weaker circumnuclear sources in NGC~253.

There are 10--15~circumnuclear sources with steep radio
spectra, likely to be supernova remnants.  The number of steep-spectrum
sources is uncertain because some are too
weak for good spectral-index determination. In addition, since
the ({\it u,v}\/) coverage at 6~cm is not matched
to that at 20~cm, more flux density may be ``resolved out''
at 6~cm, which could lead to estimates of spectra
that are overly steep.  (This resolution effect is the
primary reason that spectral indices are not quoted in 
Table~\ref{tab:sor}.)  By comparison, there
are at least 32 steep-spectrum sources in the inner 200-pc
starburst \citep{ulv97}; correcting for confusion would probably 
increase this number to 50 or more.
If the circumnuclear radio sources have the same general character
as those in the main starburst disk, a simple comparison with the
analysis of \citet{ulv97} indicates that the estimated 
circumnuclear supernova rate 
outside the central starburst is $\lesssim 0.1$~yr$^{-1}$.  (Of
course, some sources outside the strong starburst could have been
missed, due primarily to chromatic aberration.)
In any case, based solely on the ratio
of steep-spectrum sources inside and outside the 200-pc starburst, 
we infer that at least 20\%--30\% of the global star-formation and supernova
remnants are outside that central starburst.  
No supernovae have been detected in the circumnuclear
region during limited searches made at optical wavelengths in 
1988--1991 \citep{ric98}, and at near-infrared wavelengths in 
late 1993 \citep{gro99}.  However, such searches did not cover
long enough time periods to expect supernova detections in 
NGC~253, and also could have been affected by obscuration and confusion.

\begin{figure}
\figurenum{3}
\plotone{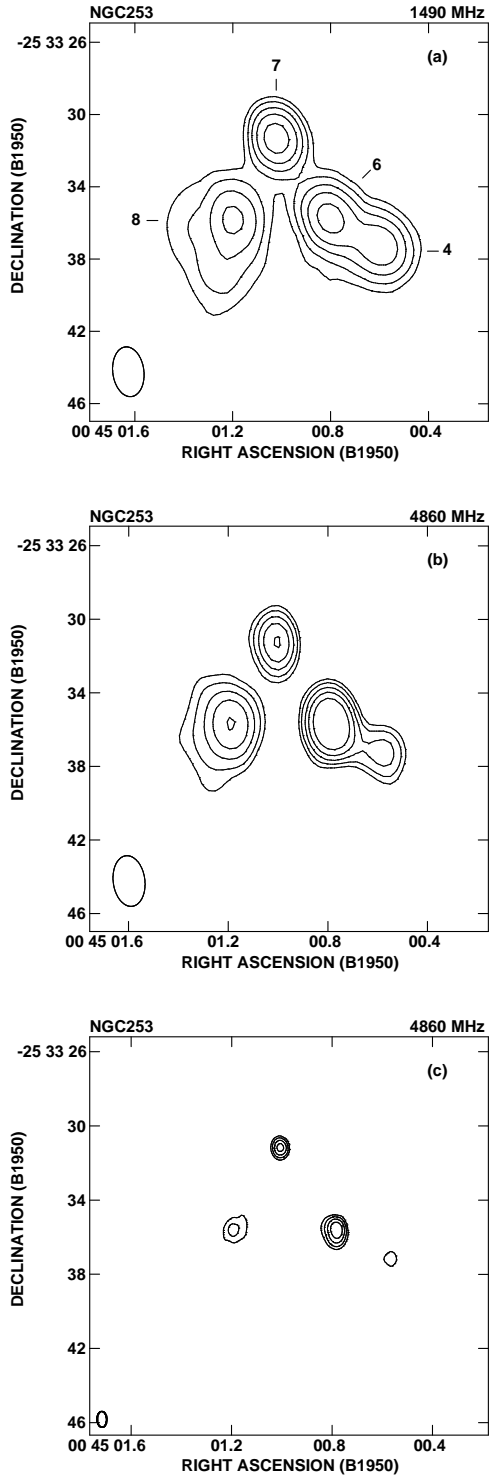}
\vspace{-3cm}
\caption{Enlargements of a complex of sources
to the west of the main starburst disk.  (a) 20-cm image, with contour
levels of $-1$, 1, 1.4, 2, 2.8, and 4 times 300~$\mu$Jy~beam$^{-1}$.
(b) 6-cm image, tapered to the same resolution as the 20-cm image, with 
contour levels of $-1$, 1, 1.4, 2, 2.8, and 4 times 200~$\mu$Jy~beam$^{-1}$.
(c) 6-cm full resolution image, with the same contour levels and restoring 
beam as in Figure~\ref{blowup}b.  \label{blowup}}
\end{figure}

Figure~\ref{blowup} shows a complex of sources to the west
of the main starburst disk, roughly 800~pc from the galaxy
nucleus.  Figure~\ref{blowup}a is the 20-cm image, showing
four individual radio sources.  These sources are numbered
4, 6, 7, and 8 in Table~\ref{tab:sor} and similarly labeled in
the figure.  Two 6-cm images also are shown. Figure~\ref{blowup}b
is an image for which the visibility data were
tapered and then restored with a point-spread function identical
to that at 20~cm, while 
Figure~\ref{blowup}c is a full-resolution image.
Inspection of Table~\ref{tab:sor} reveals that the two strongest 
6-cm sources shown in Figure~\ref{blowup}b have flat or 
inverted spectra.  Summing the four individual
compact sources gives a total 6-cm flux density of $6.6 \pm 0.4$~mJy
and a total 20-cm flux density of $8.3 \pm 0.4$~mJy, while integration 
of the images over the region of Figure~\ref{blowup} yields
a total 6-cm flux density of 7.7~mJy and a total 20-cm flux density
of 9.0~mJy, slightly higher than the fits to the compact sources.

\placefigure{blowup}

We take the somewhat simplistic step of making a two-component 
decomposition of the flux density in the radio sources shown in 
Figure~\ref{blowup}.  To do so, we assume the presence of a
thermal, flat-spectrum component having $\alpha=-0.1$, and a
nonthermal, steep-spectrum component due to supernova remnants.
The steep-spectrum
sources in the inner starburst of NGC~253 have typical
spectral indices of $\alpha=-0.7$ \citep{ulv97}, while galactic
supernova remnants have a median spectral index of $\alpha=-0.5$
\citep{gre96}, so there is some range in the estimates of the
flat-spectrum component.  The simple spectral decomposition gives
thermal radio emission containing 5.3 to 5.9~mJy out of a 
total of 6.6~mJy at 6~cm,
for a steep-spectrum component having a spectral index ranging
from $-0.5$ to $-0.7$.  Thus, 80\% to 90\% of the 6-cm flux
density in the region shown in Figure~\ref{blowup} appears
to be thermal in origin.

The natural conclusion is that
the flat-spectrum radio component is thermal radiation from \H2 regions
energized by young stars. Applying the analysis of 
the strongest flat-spectrum source in the starburst disk, which
was described by \citet{ulv97},
we find that the equivalent of $\sim 70$~O5 stars is necessary
to energize the radio emission within a region of about $12''\times 10''$
($144 \times 120$~pc).  This is comparable to the number of
young stars needed to account for the strongest thermal source 
in the inner starburst, but those stars are contained in a volume
$\sim 6\times 10^5$ times higher.  Therefore, the average thermal
gas density in the 130-pc region shown in Figure~\ref{blowup} 
is $\sim 20$~cm$^{-3}$, much
lower than the values of $\sim 10^4$~cm$^{-3}$ found in the
dense part of the inner starburst of NGC~253 \citep{ulv97} and
in NGC~5253 \citep{tur98}, or of $\sim 10^3$~cm$^{-3}$ for
the typical thermal radio sources in NGC~4038/4039 \citep{nef00}.

Few high-resolution observations at other wavelengths are available for the 
region shown in Figure~\ref{blowup}.  However, a moderate-resolution 
($9'' \times 17''$) VLA image \citep{car96} shows a source 
at the same location,
with an apparent 3.6-cm flux density between 5 and 6~mJy.  This is
entirely consistent with the value expected for
for the flat-spectrum component deduced above.  Near-infrared images 
shown in Figures~1 and 10 of 
\citet{eng98} indicate a slight enhancement in K (2.2~$\mu$m) at the
location of the complex of compact radio sources, apparently along
an inner spiral arm.  This infrared source may represent
a highly reddened set of \H2 regions containing the numerous young 
stars that energize the local radio complex.

A similar argument can be made for source 15, which lies nearly
2~kpc from the galaxy nucleus, and also appears 
to have a thermal spectrum.  Ionization of this radio source requires 
the equivalent of about 35~O5 stars in a region about 15--30~pc in diameter,
and an average ionized density near 200~cm$^{-3}$.  Chromatic
aberration makes these estimates somewhat uncertain, but they
should be correct to 50\% or better.

The relatively low average densities derived for 
the two strongest thermal-emitting complexes in the circumnuclear
region imply that they contain more ``normal'' star
formation, rather than the intense starbursts characteristic of
the central 200~parsecs.   The number and the density of massive 
young stars are significantly higher than in Orion and other nearby
Galactic O-B associations \citep{bla64}, and may be comparable
to the richest star-forming regions in our Galaxy, such as
W49 \citep{wel93,dep97}.  However, the intensity of massive
star formation is 
considerably lower than in 30~Doradus \citep{hun95} or in the super 
star clusters seen in a number of starburst galaxies
\citep{oco94,sch98,tur98,whi99,kob99,nef00}.

\section{Summary}

We have used archival VLA data to image the circumnuclear region 
of NGC~253 at arcsecond resolution.  Twenty-two compact
radio sources have been found in the inner 2~kpc of the galaxy, but 
outside the well-known 200-pc disk, and most of these 
are probably associated with regions of recent star formation.
The supernova rate inferred outside the central starburst is 
$\lesssim$0.1~yr$^{-1}$; this may be a slight underestimate
due to the decreasing sensitivity of the radio observations
at distances more than $\sim 2$\arcmin\ from the nucleus.
Therefore, the region outside the well-studied inner starburst
seems to account for a significant fraction of the recent star
formation in NGC~253.  A collection of sources located 800~pc 
to the west of the nucleus appears to be a complex of \H2 
regions energized by the equivalent of 70~O5 stars, but with 
an average ionized gas density $\sim 10^3$ times lower than 
that found in the inner starburst of the galaxy.

\acknowledgments

I thank K. Anantharamaiah, R. Antonucci, N. Mohan, and W. Pietsch for 
useful discussions and for providing data in advance of publication.
I especially thank the referee, Jean Turner, for some perceptive
comments and for pointing out errors in Table~\ref{tab:sor}.
This research has made use of the NASA/IPAC Extragalactic Database
(NED) which is operated by the Jet Propulsion Laboratory, California
Institute of Technology, under contract with the National 
Aeronautics and Space Administration.  In addition, this research
has made use of NASA's Astrophysics Data System.

\clearpage

\begin{deluxetable}{lccccc}
\tablecolumns{6}
\tablewidth{0pc}
\tablecaption{Full-Resolution Image Parameters}
\tablehead{
\colhead{$\lambda$} & \colhead{Date} & \multicolumn{3}{c}{Beam Size} &
\colhead{r.m.s. noise} \\
\cline{3-5} \\
&&\colhead{Major} & \colhead{Minor} & \colhead {P.A.} & \\
\colhead{(cm)} && \colhead{(\arcsec)} & \colhead{(\arcsec)} 
& \colhead{(\arcdeg)} & \colhead{($\mu$Jy beam$^{-1}$)} }
\startdata
6&87JUL10&0.86&0.51&2&27 \\
20&87JUL21&2.77&1.72&7&41 \\
\enddata
\label{tab:obssum}
\end{deluxetable}

\begin{deluxetable}{lllccc}
\tablecolumns{7}
\tablewidth{0pc}
\tablecaption{Compact Circumnuclear Sources}
\tablehead{
\colhead{No.}&\multicolumn{2}{c}{Position}&\multicolumn{2}{c}{Flux Density (mJy)}&
\colhead{Source Size\tablenotemark{a}} \\
\cline{2-3} \cline{4-5} \\
&\colhead{$\alpha(1950)$} & \colhead{$\delta(1950)$} & \colhead{6 cm} &
\colhead{20 cm} & \colhead{(arcsec)} }
\startdata
\ 1&$00^h44^m59.\!^s 404$ &$-25^\circ 33'59.\!''07$ & $0.4\pm 0.1$ & $0.6\pm 0.1$ &  U \\
\ 2&00 44 59.498 &$-$25 33 54.91  & $0.3\pm 0.1$ & $0.7\pm 0.1$ &  U \\
\ 3&00 45 00.117 &$-$25 34 20.89  & $0.2\pm 0.1$ & $0.6\pm 0.1$ &  U \\
\ 4\tablenotemark{b}&00 45 00.580 &$-$25 33 37.30  & $0.7\pm 0.2$ & $2.3 \pm 0.2$& 0.9 $\times$ 0.7, 89\arcdeg \\ 
\ 5&00 45 00.619 &$-$25 34 15.25  & $<0.2$ & $0.4\pm 0.1$ & U \\
\ 6&00 45 00.785 &$-$25 33 35.66  & $2.4\pm 0.2$ & $2.8\pm 0.2$ & 1.1 $\times$ 0.9, 7\arcdeg \\
\ 7&00 45 01.007 &$-$25 33 31.19  & $1.2\pm 0.1$ & $1.8\pm 0.2$ & 0.6 $\times$ 0.6 \\
\ 8&00 45 01.187 &$-$25 33 35.61  & $2.3\pm 0.2$ & $1.4\pm 0.1$ & 1.9 $\times$ 1.5, 145\arcdeg \\
\ 9&00 45 02.495 &$-$25 34 00.40  & $1.2\pm 0.1$ & $3.5\pm 0.2$ & U \\
10&00 45 02.659 &$-$25 33 04.87  & $<0.2$ & $0.4\pm 0.1$ & U \\
11&00 45 03.346 &$-$25 32 59.96  & $0.4\pm 0.1$ & $0.9\pm 0.1$ & U \\
12&00 45 03.497 &$-$25 33 00.15  & $<0.2$ & $0.6\pm 0.1$ & U \\
13&00 45 04.236 &$-$25 32 46.97  & $0.3\pm 0.1$ & $0.5\pm 0.1$ &      U \\
14&00 45 05.998 &$-$25 32 29.16  & $0.2\pm 0.1$ & $0.4\pm 0.1$ &      U \\
15\tablenotemark{c}&00 45 06.097 &$-$25 31 55.60  & $2.7\pm 0.3$ & $2.4\pm 0.3$ & \nodata \\
16&00 45 06.153 &$-$25 32 26.66  & $0.5\pm 0.1$ & $0.6\pm 0.1$ &      U \\
17&00 45 09.602 &$-$25 33 55.91  & $<0.2$ & $0.5\pm 0.1$  &      U \\
18\tablenotemark{b}&00 45 09.804 &$-$25 33 06.73  & $1.5\pm 0.1$ & $3.9\pm 0.2$ & 1.2 x 0.9, 2\arcdeg \\
19\tablenotemark{c}&00 45 12.084 &$-$25 32 56.00  & $0.4\pm 0.2$ & $2.4\pm 0.3$ & \nodata \\
20\tablenotemark{c}&00 45 13.136 &$-$25 33 00.97  & $2.7\pm 0.2$ & $6.1\pm 0.3$ & \nodata \\
21&00 45 13.308 &$-$25 33 08.83  & $<0.2$ & $0.6\pm 0.1$ & U \\
22\tablenotemark{c}&00 45 15.846 &$-$25 32 01.53  & $0.9\pm 0.2$ & $1.2\pm 0.2$ & \nodata \\
\enddata
\label{tab:sor}
\tablenotetext{a}{Source sizes denoted as ``U'' are unresolved.}
\tablenotetext{b}{Source is significantly larger at 20 cm
than at 6 cm.  Derived spectral index may be too steep, because of resolution
effects at 6 cm.}
\tablenotetext{c}{Source has significant primary beam correction 
and chromatic aberration, causing uncertainties in the source
flux densities and structure.  Only the total flux densities have 
been measured, and the 6-cm flux density may be an underestimate.}
\end{deluxetable}
\clearpage

\begin{deluxetable}{lllccl}
\tablecolumns{7}
\tablewidth{0pc}
\tablecaption{Wide-Field Radio Sources\tablenotemark{a}}
\tablehead{
\colhead{No.}&\multicolumn{2}{c}{Position}&\multicolumn{2}{c}{Flux Density (mJy)}&
\colhead{Comments} \\
\cline{2-3} \cline{4-5} \\
&\colhead{$\alpha(1950)$} & \colhead{$\delta(1950)$} & \colhead{6 cm} &
\colhead{20 cm} & }
\startdata
B1&$00^h44^m45.\!^s 07$ &$-25^\circ 34'05.\!''5$ & \nodata & $16\pm 3$ & Outside 6 cm window \\
B2a&00 45 17.08 &$-$25 29 58.6  & $< 2$ & $5\pm 1$ & Undetected at 6 cm\\
B2b&00 45 17.45 &$-$25 29 59.7 &   $1.7\pm 0.3$ & \nodata & Blends with B2c at 20 cm\\
B2c&00 45 17.54 &$-$25 30 00.6  &  $5.2\pm 0.8$ & $18\pm 3$ & Blends with B2b at 20 cm \\
B3&00 45 31.33 &$-$25 34 52.6  & \nodata & $50\pm 8$ & Outside 6 cm window \\
\enddata
\label{tab:back}
\tablenotetext{a}{Source sizes are unknown, due to severe chromatic aberration.}
\end{deluxetable}
\clearpage

\clearpage

\clearpage

\clearpage

\end{document}